\documentclass[prl,aps,reprint,superscriptaddress]{revtex4-1} %  preprint  

\usepackage{amsfonts}
\usepackage{amsmath}
\usepackage{amssymb}
\usepackage{mathrsfs}

\usepackage[latin9]{inputenc}
\usepackage[T1]{fontenc}
\usepackage[french,english]{babel}
\usepackage{graphicx}
\usepackage{float}
\usepackage{amssymb}
\usepackage{amsmath}
\usepackage{epstopdf}
\usepackage{xfrac}
\usepackage{bm}
\usepackage{esint}
\usepackage{lipsum}
\usepackage{color}

\providecommand\bcdot{\boldsymbol{\cdot}}

\begin{document}

\title[]{
Asymptotic network models of subwavelength metamaterials \\ formed by closely  packed
 photonic and phononic crystals}
\author{Alice L. Vanel, Ory Schnitzer, Richard V. Craster}
\affiliation{Department of Mathematics, Imperial College London, London SW7 2AZ, UK}

\begin{abstract}
We demonstrate that photonic and phononic crystals consisting of closely spaced inclusions constitute a versatile class of subwavelength metamaterials. Intuitively, the voids and narrow gaps that characterise the crystal form an interconnected network of Helmholtz-like resonators. We use this intuition to argue that these continuous photonic (phononic) crystals are in fact asymptotically equivalent, at low frequencies, to discrete capacitor-inductor (mass-spring) networks whose lumped parameters we derive explicitly. The crystals are tantamount to metamaterials as their entire acoustic branch, or branches when the discrete analogue is polyatomic, is squeezed into a subwavelength regime where the ratio of wavelength to period scales like the ratio of period to gap width raised to the power $1/4$; at yet larger wavelengths we accordingly find a comparably large effective refractive index. The fully analytical dispersion relations predicted by the discrete models yield dispersion curves that agree with those from finite-element simulations of the continuous crystals. The insight gained from the network approach is used to show that, surprisingly,  the continuum created by a closely packed hexagonal lattice of cylinders is represented by a discrete honeycomb lattice. The analogy is utilised to show that the hexagonal continuum lattice has a  Dirac-point degeneracy that is lifted in a controlled manner by specifying the area of a symmetry-breaking defect. 
\end{abstract}

\maketitle

\textit{Introduction}.---Mechanical mass-spring networks and electromagnetic circuit models have long acted to motivate, and gain qualitative intuition, in the study of  waves in solid-state physics \cite{brillouin53a}, continuous media containing periodic arrays of inclusions such as photonic and phononic crystals \cite{Sievenpiper:98,*Staffaroni:12}, and more recently in nanophotonics \cite{Engheta:05,*Engheta:07,*Alu:11}, metasurfaces \cite{Sievenpiper:99} and metamaterials \cite{pendry99a}. Typically the analogy between the continuous model and its discrete analogy is neither exact nor is it believed (let alone shown) to represent a systematic approximation. Rather, such analogies are most often introduced heuristically to aid interpretation and the lumped parameters are estimated and accepted as qualitative. In this Letter we develop asymptotically exact network analogies for two-dimensional photonic, or phononic, crystals formed by closely spaced metallic or rigid cylinders, respectively, and show that such materials constitute a versatile and tuneable family of subwavelength metamaterials.  
 
The analogies are developed, sequentially in the acoustic and electromagnetic cases, from first principles, leading to an intuitive physical description where the gaps and voids separating the closely spaced inclusions form a coupled network of Helmholtz- or LC-like resonators. The equivalent networks, whose lumped parameters are obtained explicitly, have a low cut-off frequency, effectively squeezing the entire acoustic branch into the subwavelength regime. With the discrete network at hand we can easily predict how to tune the crystal to possess desirable physical properties. In particular, we show that at frequencies well below cut-off  the closely packed crystals have a large effective refractive index, and extract its value in closed form. The discrete system also allows accessing problems of topical interest in a simplified setting, with conclusions fed back to the continuum structure. As an example we show that a closely packed hexagonal array of cylinders is analogous to a discrete {\it honeycomb} system and so, surprisingly, exhibits a Dirac point. Upon introducing a geometric perturbation at the continuum level, the equivalent asymptotic network becomes diatomic and asymmetric, having the same form known to exhibit the topological valley-Hall effect \cite{Pal:17}. 

{\textit{Phononic crystals.---}}
\label{sec:intuitive}
Consider first the acoustic scenario of a phononic crystal of rigid cylinders surrounded by an ideal gas. To generate an intuitive description of the physics, in the limit where the inclusions are closely spaced, we draw upon the concepts that underlie classical models of the Helmholtz resonator \cite{rayleigh1870}.  The resonator physics is dominated by the motion of a well defined slug of gas that moves back and forth in the mouth of the resonator, in response to external pressure changes, resisted by the compressible gas within the resonator. Similarly, here we picture the continuous phononic crystal as a network of bubbly voids inter-connected by narrow gaps in which incompressible slugs of gas oscillate in response to the pressure difference between the neighbouring voids. 

\begin{figure}
\centering
\includegraphics[scale=0.32]{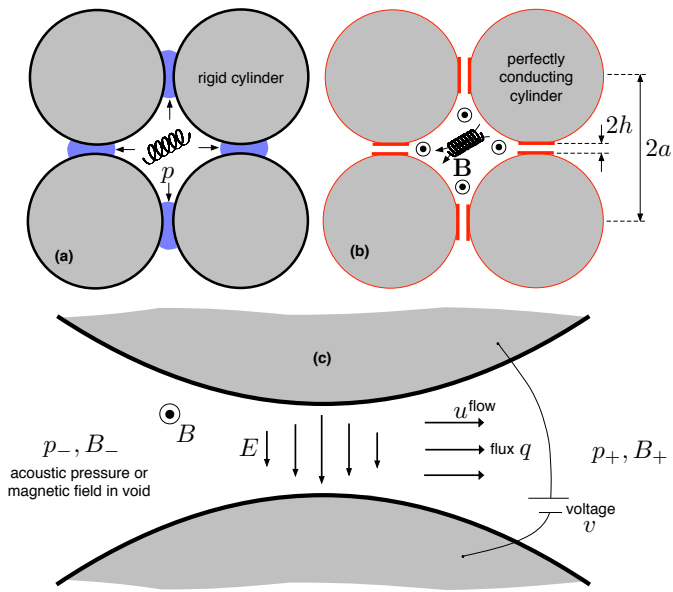}
\caption{Closely packed phononic (a) and photonic (b) crystals. In the limit $h/a\to0$, and at low frequencies, the narrow parabolic gaps act as incompressible slugs of gas in the phononic case and as parabolic capacitors in the photonic case (c). The fluid in the phononic voids behaves as a uniform ideal gas undergoing adiabatic compression (a) while the photonic voids store magnetic energy induced by currents flowing along the cylindrical surfaces (b). } 
\label{fig:figIntro}
\end{figure}
Although the theory is not limited to a specific lattice or cylinder geometry, it is didactic to first consider the simplest configuration of a square lattice of closely spaced circular inclusions, as shown in Fig.~\ref{fig:figIntro}(a); the spacing between the inclusions, $2h$, is small compared with the pitch, $2a$. Since the bubbly voids are bounded by rigid boundaries, at low frequencies the pressure $p$ in any such void 
is approximately uniform and satisfies the adiabatic relation $pA^\gamma=p_0A_0^\gamma$, where $A$ and $\gamma$ are the area of the void and the ratio of specific heats, respectively, and $p_0,A_0$ are reference values. Placing this void under compression so that the pressure is
perturbed to $p_0+\Delta p$ reduces the area of gas to $A_0-\Delta A$. If $Q$ denotes the net area flux into the void, associated with the total inward motion of the surrounding gas slugs, then $\Delta A=Q \Delta t$.  Hence, moving from the language of increments to that of calculus, the rate of change of pressure within the void is
\begin{equation}\label{adiabatic}
\frac{\partial p}{\partial t}= \frac{\gamma p_0}{A_0}Q.
\end{equation}

In general the pressure may vary from void to void, implying rapid pressure variations across the gaps. Conservation of momentum relates these pressure gradients to acceleration,
\begin{equation}\label{momentum}
\rho_0\frac{\partial {\bf u}}{\partial t}=-\nabla p,
\end{equation}
where $\bf u$ is the fluid's velocity field. 
Consider a single representative gap, like the one shown in Fig.~\ref{fig:figIntro}(c): The gap is narrow having slowly varying  boundaries that can be locally approximated by parabolas. Accordingly, the flow through the gap is approximately incompressible, uni-directional and transversely uniform; if $x$ denotes an axial coordinate symmetric about the middle of the gap, and $u$ and $q$ the velocity and total flux in that direction, respectively, then
\begin{equation}\label{eq:q}
q(t) = 2\left(h+\frac{x^2}{2a}\right)u(x,t).
\end{equation}
Thus by integrating \eqref{momentum} along the axis of the gap we find the pressure jump between the voids,
\begin{equation}\label{Newton}
p_+-p_-=-\frac{\rho_0}{\delta}\frac{dq}{d t},
\end{equation}
where $\delta$ is a normalised acoustic conductivity defined as 
\begin{equation}\label{kformula}
\delta=\frac{1}{\pi}\sqrt{\frac{2h}{a}}\ll1.
\end{equation}

Eqs.~\eqref{adiabatic} and \eqref{Newton} provide a systematic mechanical interpretation of the voids and gaps of the closely spaced phononic crystal, which is entirely explicit and remarkably does not involve any fitting or fudge parameters. In particular, \eqref{Newton} is nothing but Newton's second law applied to the slug of gas in a representative gap, the forcing being provided by the pressure jump between the neighbouring voids. Note that $\delta$ is a small parameter, meaning that the fluids constrained within the narrow gaps specifically act as heavy masses, which is consistent with our anticipation that the crystal behaves as a subwavelength metamaterial. The mechanical network is closed by \eqref{adiabatic}, which gives the constitutive law of the bubbly voids. The latter act as generalised springs, where the pressure perturbation is proportional to the accumulated sum of the inward displacements of the surrounding effective masses. Before proceeding to study the discrete media implied by \eqref{adiabatic} and \eqref{Newton}, we digress to discuss the analogous equations in the electromagnetic case. 

{\textit{Photonic crystals.---}}
The electromagnetic analogy emerges from considering an array of perfectly
conducting cylinders embedded in a dielectric matrix, in the case where the magnetic field $\bf{B}$ is polarised along the cylinders 
and the electric field ${\bf E}$ lies within the plane.
The gap in Fig.~\ref{fig:figIntro}(c) now consists of a thin vacuum %dielectric 
layer sandwiched between two
locally parabolic perfect conductors. This arrangement acts as a capacitor, %quasi-static 
where the electric field in the gap is dominated by the transverse electric-field component $E(x,t)$, where $x$ is defined as before. 
Thus the analogue of \eqref{eq:q}, 
\begin{equation}
v(t) = 2\left(h+\frac{x^2}{2a}\right) E(x,t),
\label{eq:E}
\end{equation}
 relates the voltage across the gap to the electric field, 
where the signs of $v$ and $E$ are as shown in Fig.~\ref{fig:figIntro}(c).
We now consider Amp\`ere's law in integral form, 
\begin{equation}
 \int {\bf B}\cdot d{\bf l}=\mu\epsilon {\frac{d}{dt}}\int {\bf E} \cdot d{\bf S},
\end{equation}
and consider it in a plane tangent to the axis of the cylinders and
 bisecting the gap; using a closed loop in that plane, that enters the
 voids, the analogue of \eqref{Newton}
 emerges as
\begin{equation}
  B_+-B_- =\frac{\mu\epsilon}{\delta} \frac{dv}{dt},
\label{eq:jump2}
\end{equation}
where referring to Fig.~\ref{fig:figIntro}(c) the magnetic-field component $B$ is defined positive in the direction pointing away from the page. Noting that the charge per-unit-length in the gap is $\epsilon v/ \delta$, we see that the capacitance per-unit-length of the gap is {$\epsilon/\delta$}. 

Time variations of the magnetic field in the voids are resisted by inductance. Thus, according to Faraday's law, % of induction,  
\begin{equation}
\oint {\bf E}\cdot d{\bf l}=-\frac{d}{dt}\int {\bf
  B}\cdot{\bf n}dS,
\label{eq:faraday0}
\end{equation}
the time-variation of the magnetic flux through any given void is proportional to the voltage in a loop surrounding it. Since the inclusions are perfectly conducting then, at low frequencies, the magnetic field is uniform in the void, and also the total voltage corresponds to the sum of voltages over the gaps surrounding it.  Hence 
\begin{equation}
V=A_0\frac{dB}{dt},
\label{eq:faraday}
\end{equation}
where 
 the total loop voltage $V$ is positive counter-clockwise. Eqs.~\eqref{eq:jump2} and \eqref{eq:faraday} are analogous to \eqref{Newton} and \eqref{adiabatic}, respectively, and provide an explicit electromagnetic interpretation of the continuous photonic crystal as a wired network of capacitors where the currents sloshing charge in between the gap capacitors are resisted by inductance of the coupled circuit loops. 

{\textit{Network equations.---}}
\label{sec:network}
The lattice structure is now recreated by labelling the voids using the indices $n,m$ and denoting the pressure in, and net flux into, void $n,m$ by $p_{n,m}$ and $Q_{n,m}$, respectively. By summing \eqref{Newton} applied to the gaps surrounding void $n,m$ we find 
\begin{multline}\label{Qnet}
\frac{\rho_0}{\delta}\frac{dQ_{n,m}}{dt}
= p_{n,m+1}+p_{n,m-1}+p_{n-1,m}\\+p_{n+1,m}-4p_{n,m},
\end{multline}
thence from \eqref{adiabatic} a discrete wave equation emerges that
distills all the relevant physics into a concise form
\begin{multline}
\frac{A_0}{c^2\delta}
\frac{d^2p_{n,m}}{d t^2} = p_{n,m+1}+p_{n,m-1}+p_{n-1,m}\\ +p_{n+1,m}-4p_{n,m},
\label{eq:mass-spring}
\end{multline}
where $c^2=\gamma p_0/\rho$ is the speed of sound squared. A similar equation is found in the electromagnetic equation  for the magnetic field $B_{n,m}$ in void ${n,m}$, with $c^2=1/(\mu\epsilon)$ now denoting the speed of light. This equation is found from \eqref{eq:faraday} once the analog of \eqref{Qnet} is obtained from \eqref{eq:jump2}, with $\rho_0 Q_{n,m}$ replaced by $\mu \epsilon V_{n,m}$, where $V_{n,m}$ is the voltage over a counter-clockwise loop around void ${n,m}$.  

The analysis given for a square lattice of circular cylinders of radius $a-h$, with $h\ll a$, readily
generalises to an arbitrary two-dimensional lattice --- of arbitrarily shaped inclusions and not necessarily periodic ---  as long as the inclusions are separated by narrow and locally circular gaps. The discrete wave equation is derived as before by applying \eqref{adiabatic} and \eqref{Newton} (or their electromagnetic analogues) to the voids and gaps, in general allowing the area $A_0$ of each void and geometric parameter $\delta$ of each gap to differ; the latter parameter is still given by \eqref{kformula}, but with $1/a$ replaced by $(1/R_1+1/R_2)/2$, where $R_1$ and $R_2$ are the local radii of curvature of the inclusion surfaces bounding the gap. 

It is also worth mentioning that the above simplification of a continuous periodic medium of closely packed inclusions applies to any generic wave field $p$ governed by the wave equation $\partial^2p/\partial^2 t -c^2\nabla^2 p =0$, where $c$ is the relevant wave speed and $p$ satisfies Neumann conditions $\partial{p}/\partial{n}=0$ on inclusion boundaries. In fact, the heuristic presentation here can be formalised using the method of matched asymptotic expansions \cite{Crighton:12}, which confirms that the reduced network models constitute a precise asymptotic approximation in the limit where $h/a\to0$ and the wavelength is large compared to the pitch. 

\begin{figure}
\centering
\includegraphics[scale=0.28]{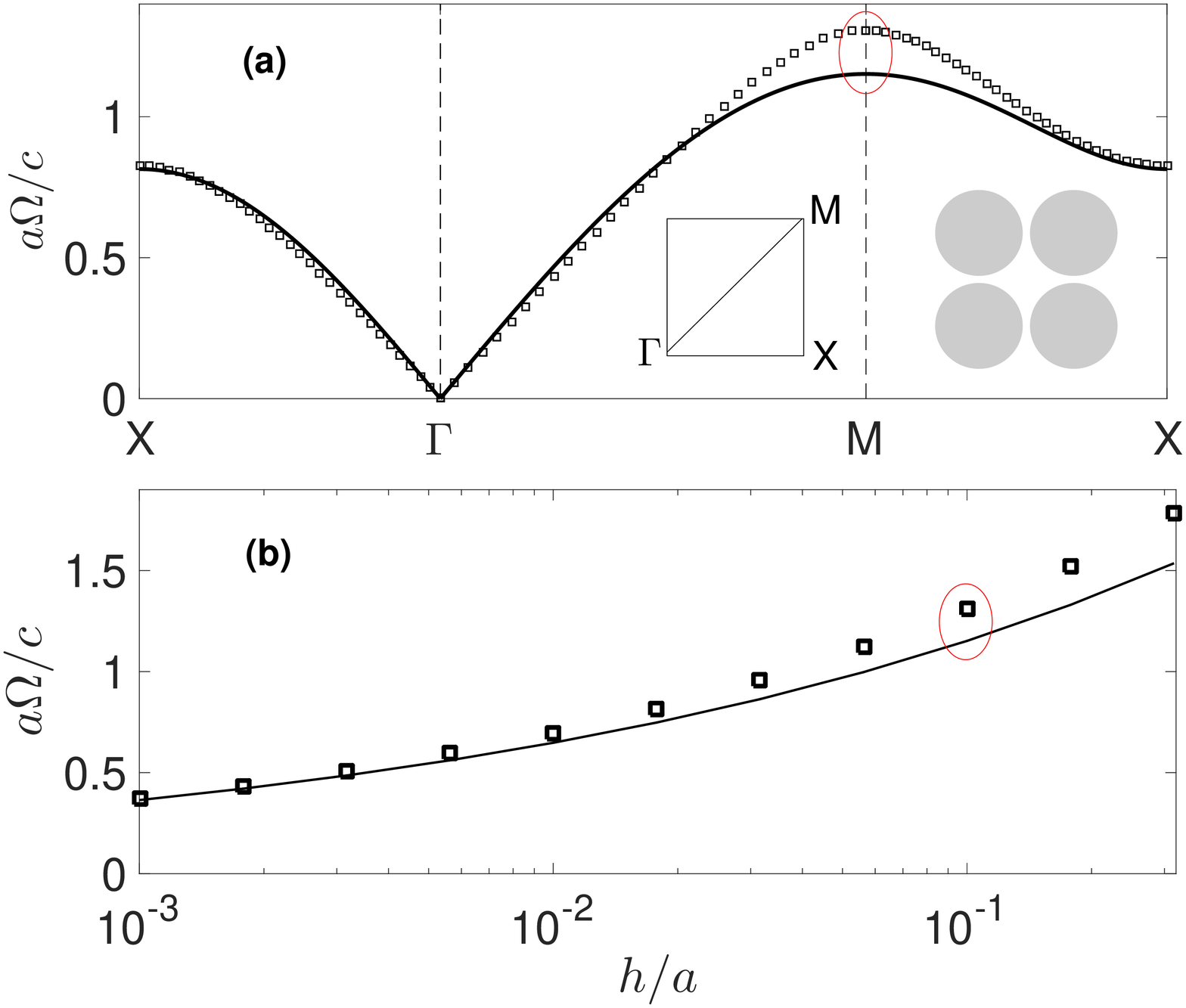}
\caption{Panel (a) shows the acoustic branch for a square array of
  cylinders, pitch $2a$, for $h/a=0.1$. The dispersion curves are obtained
  numerically (symbols) and from the approximate formula 
  \eqref{eq:general} (solid lines). 
Panel (b) compares the numerical and approximate predictions at point $M$ of the irreducible Brillouin zone, demonstrating that the latter constitutes a precise asymptotic approximation in the limit $h/a\to0$. The points highlighted in red act to connect the data from one panel to the other.}
\label{fig:square}
\end{figure}
\textit{Acoustic branch.---} Normally, phononic and photonic crystals are non-dispersive in the subwavelength regime, and therefore can be described by averaged quasi-static properties. Here, in contrast, the narrowness of the gaps gives rise to dispersion and band gaps in that regime, suggesting that closely spaced crystals can function as metamaterials. In particular, substituting into \eqref{eq:mass-spring} a Bloch-wave ansatz $p_{n,m}=\exp[i(\boldsymbol{\kappa}\bcdot\mathbf{x}_{n,m}-\Omega t)]$, where $\boldsymbol{\kappa}=(\kappa_x,\kappa_y)$ is the Bloch wavevector and $\mathbf{x}_{n,m}$ marks the position of void ${n,m}$, we find
\begin{equation}
\Omega^2=\frac{2c^2\delta}{
A_0}\left[2-\cos(2\kappa_xa)-\cos(2\kappa_ya)\right].
\label{eq:general}
\end{equation}
This dispersion relation corresponds to the acoustic branch of the closely packed square lattice of cylinders, which, noting the dependence of $\delta$ upon $h/a$, is squeezed into a subwavelength regime where the wavelength and period scale at a ratio of $(a/h)^{1/4}$ to one. (The next branch of the closely packed crystal lies  at higher frequencies where wavelength is comparable to the pitch, hence there is a large band gap above the acoustic branch.) Fig. \ref{fig:square}(a) compares the asymptotic dispersion relation \eqref{eq:general} in the case of circular inclusions, in which case $A_0=(4-\pi)a^2$, with full finite element numerical
simulations. We choose $h/a=0.1$ to show that the agreement is gratifying even for moderately small 
values of $h/a$. As demonstrated in Fig. \ref{fig:square}(b), which shows the maximum of the acoustic branch as a function of $h/a$, the dispersion relation \eqref{eq:general} is asymptotic as $h/a\to0$. 

\begin{figure}[b]
      \includegraphics[scale=0.5]{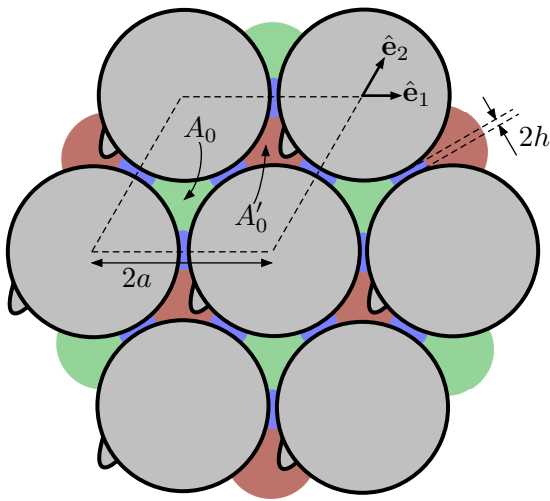}
\caption{A closely packed hexagonal lattice of cylinders forms a honeycomb network of voids connected by narrow gaps. Inversion symmetry is broken here by making the areas of the two voids in each unit cell different.}
\label{fig:defect}
\end{figure}

\textit{Hexagonal lattice with a Dirac point.---}
\label{sec:dirac}
The physical interpretation of the gaps and voids as the elements forming the equivalent network leads to surprising predictions. For instance, a natural assumption is that a honeycomb array of cylinders will mimic the arrangement of graphene and thence allow for Dirac points in the dispersion surfaces. As shown in Fig.~\ref{fig:defect}, however, it is the
hexagonal array of cylinders that produces the Dirac point,
the six voids surrounding the central cylinder forming a honeycomb array.
In the case where the cylinders are perfectly circular then there are two identical voids in each unit cell, of area $A_0=(\sqrt{3}-\pi/2)a^2$. More generally, we allow the area of the two voids
to differ, say due to a defect such as that shown in Fig.~\ref{fig:defect}. We then find the coupled 
network equations, 
\begin{gather}\label{honey network}
\frac{A_0}{c^2\delta}\frac{\partial^2p_{n,m}}{\partial t^2}=p'_{n,m}+p'_{n-1,m}+p'_{n,m-1}-3p_{n,m},\\
\frac{A_0'}{c^2\delta}\frac{\partial^2p'_{n,m}}{\partial t^2}=p_{n,m}+p_{n+1,m}+p_{n,m+1}-3p'_{n,m},
\end{gather}
where $p_{n,m}$ and $p'_{n,m}$ are the pressures in the two voids, respectively of areas $A_0$ and $A_0'$, within unit cell $(n,m)$; the indexes $n$ and $m$ represent $2a$ displacements in the $\hat{\mathbf{e}}_1=\hat{\mathbf{e}}_x$ and $\hat{\mathbf{e}}_2$ directions shown in Fig.~\ref{fig:defect}. The dispersion relation follows as
\begin{multline}\label{honeydisp}
\frac{A_0A_0'}{c^4\delta^2}\Omega^4-\frac{3(A_0+A_0')}{c^2\delta}\Omega^2+6 -2\cos(2\kappa_x a) \\ -4\cos(\kappa_x a)\cos(\sqrt{3}\kappa_y a)=0.
\end{multline}
\begin{figure}[t]
   \includegraphics[scale=0.28]{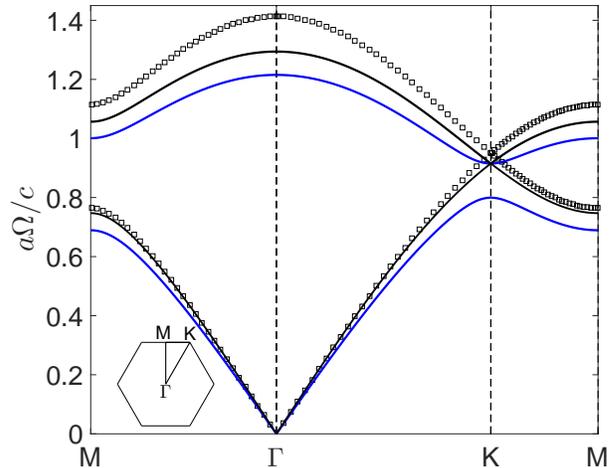}
	\caption{Dispersion curves for the acoustic branch for the hexagonal
  array of circular cylinders ($h/a=0.01$, symbols are numerics and black line is \eqref{honeydisp} for $A_0=A_0'$). Insertion of a defect opens the hexagon-lattice Dirac point %in the hexagon lattice 
  at $K$ (solid blue line is \eqref{honeydisp} for $A_0'-A_0=0.05a^2$).}
\label{fig:hexagon}
\end{figure}

Fig.~\ref{fig:hexagon} depicts the predictions of the network approximation \eqref{honeydisp} for $h/a=0.01$. In the degenerate case, where $A_0=A_0'$,  the discrete theory predicts a Dirac cone at the $K$ point, where $\Omega=\Omega_D=c\sqrt{3\delta/A_0}$. This prediction agrees with full-wave finite-element simulations of the original \emph{hexagonal} arrangement of closely spaced cylinders. We also show the predictions of \eqref{honeydisp} in the case where a defect is introduced such that $A_0'\ne A_0$; a band gap opens where the Dirac point used to be and the frequencies at the $K$ point become $\Omega_D$ and $\Omega'$, where
\begin{equation}
\frac{\Omega'-\Omega_D}{\Omega_D} = \sqrt{\frac{A_0}{A_0'}}-1.
\end{equation}
Thus our theoretical approach reveals that a closely packed hexagonal lattice of cylinders possesses a Dirac degeneracy, which can be removed in a controlled fashion by modifying a lumped geometric parameter. In particular, the Dirac degeneracy is removed here by breaking inversion symmetry, in which case the crystal can be designed to exhibit the topological valley-Hall effect \cite{Xiao:07,Ma:16}. In fact, the reduced network model \eqref{honey network} of the closely packed hexagonal lattice is identical to the diatomic mass-spring honeycomb lattice studied in Ref.~\cite{Pal:17}, where this effect is explored in detail. 

 \begin{figure}[t]
   \includegraphics[scale=0.45]{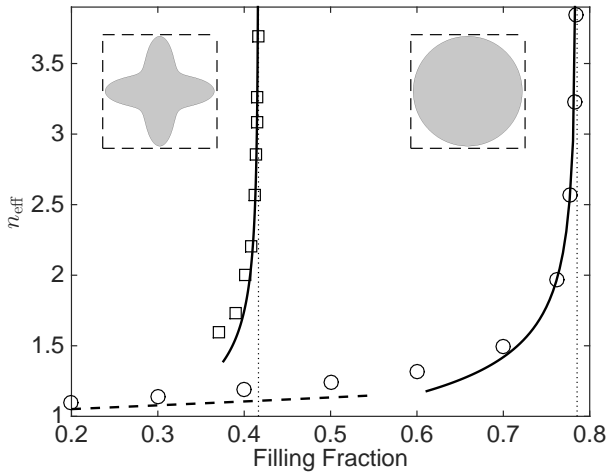}
\caption{Effective refractive index $n_{\text{eff}}$ at long wavelengths versus the filling fraction of a square lattice of circular and leaf-shaped inclusions. (The leaf shape is $\rho+\rho\sin^2({2\theta})=a(1-h)$ in polar coordinates $(\rho,\theta)$, for which $R_1=R_2=a/9$ and $A_0/a^2=4-{3\pi}/{4\sqrt{2}}$.) Solid lines: closely spaced limit \eqref{eq:neff}; dashed line:  Maxwell Garnett formula \cite{maxwellgarnett04a} for the dilute limit (circular inclusions); symbols: numerical simulations. }
\label{fig:neff}
\end{figure}
{\textit{High-index metamaterials.---}}
\label{sec:effective}
In the long-wavelength limit, namely when the ratio of the wavelength to period is large compared with $(a/h)^{1/4}$, the frequency is well below the subwavelength Helmholtz-type (or LC-type) resonances of the network. Accordingly, the dispersion surface is conical and at large scales the material effectively acts as a homogeneous transparent medium characterised by a large refractive index $n_{\text{eff}}$. For example, the effective index for a square lattice is read off Eq. \eqref{eq:general}, 
\begin{equation}
 n_{\text{eff}}=\frac{1}{2}\sqrt{\frac{A_0}{a^2\delta}},
 \label{eq:neff}
\end{equation}
which  scales with the gap width as $(h/a)^{-{1}/{4}}$ [cf.~\eqref{kformula}]; our prediction for the effective index is corroborated with numerical simulations in Fig.~\ref{fig:neff}, for circular and leaf-shaped inclusions.  While our theory appears to provide a recipe for a material with an arbitrarily high refractive index, clearly there are upper bounds owing to manufacturing limitations and physics disregarded in our analysis. Thus an important extension to this work would be to relax the assumptions that the gas is ideal and that the inclusions are perfectly rigid in the acoustics case, and that the inclusions are perfect conductors in the electromagnetic case.  

{\textit{Concluding remarks.---}}
\label{sec:conclude}
We were drawn to the intuitive framework developed in this Letter whilst asymptotically modelling closely spaced phononic and photonic crystals. Thus, whereas asymptotic limiting descriptions exist, and are of considerable value, in the dilute limit \cite{nicorovici95b},  in this equally important concentrated limit a versatile approximation method was lacking. Yet this is precisely where standard semi-analytical methods, such as plane-wave \cite{joannopoulos08a} and multi-pole \cite{movchan02a,nicorovici95b} expansions, as well as direct numerical methods naturally struggle. We here obtained asymptotic network approximations in that limit, from which we extracted generic scalings and  formulae revealing, in hindsight, that closely spaced crystals can be tuned to operate as subwavelength metamaterials. In particular, this class of metamaterials can be viewed as a two-dimensional idealisation, also generalised to acoustics, of the three-dimensional strongly capacitive metallic structures introduced by Sievenpiper \textit{et al.} \cite{Sievenpiper:98}, which have similar properties. This philosophy of exploiting the narrowness of gaps, which is now supplemented by a simple analytical framework, has broad application across both electromagnetics and acoustics.

R. V. Craster thanks the EPSRC for their support through grant {EP/L024926/1}. 

\bibliographystyle{apsrev4-1}
\bibliography{references}

\end{document}